\documentclass[a4paper,11pt]{article}
\pdfoutput=1 % if your are submitting a pdflatex (i.e. if you have
\usepackage{jcappub} % for details on the use of the package, please
\usepackage{indentfirst}
\usepackage{appendix}
\usepackage{multirow}

\usepackage{array}
\usepackage{enumerate}
\usepackage{graphicx}
\usepackage{dcolumn}
\usepackage{bm}
\usepackage{amssymb}
\usepackage{latexsym}
\usepackage{booktabs}
\usepackage{amsmath}
\usepackage{multirow}
\usepackage{url}

\begin{document}

\title{Prospects for measuring the Hubble constant and dark energy using gravitational-wave dark sirens with neutron star tidal deformation}

\author[a]{Shang-Jie Jin,}
\author[a]{Tian-Nuo Li,}
\author[a]{Jing-Fei Zhang}
\author[a,b,c,1]{and Xin Zhang\note{Corresponding author.}}
\affiliation[a]{Key Laboratory of Cosmology and Astrophysics (Liaoning) and College of Sciences, Northeastern University, Shenyang 110819, China}
\affiliation[b]{Key Laboratory of Data Analytics and Optimization for Smart Industry (Ministry of Education), Northeastern University, Shenyang 110819, China}
\affiliation[c]{National Frontiers Science Center for Industrial Intelligence and Systems Optimization, Northeastern University, Shenyang 110819, China}

\emailAdd{jinshangjie@stumail.neu.edu.cn, litiannuo@stumail.neu.edu.cn, jfzhang@mail.neu.edu.cn, zhangxin@mail.neu.edu.cn}

\abstract{Using the measurements of tidal deformation in the binary neutron star (BNS) coalescences can obtain the information of redshifts of gravitational wave (GW) sources, and thus actually the cosmic expansion history can be investigated using solely such GW dark sirens. To do this, the key is to get a large number of accurate GW data, which can be achieved with the third-generation (3G) GW detectors. Here we wish to offer an answer to the question of whether the Hubble constant and the equation of state (EoS) of dark energy can be precisely measured using solely GW dark sirens. We find that in the era of 3G GW detectors ${\cal O}(10^5-10^6)$ dark siren data (with the NS tidal measurements) could be obtained in three-year observation if the EoS of NS is perfectly known, and thus using only dark sirens can actually achieve the precision cosmology. Based on a network of 3G detectors, we obtain the constraint precisions of $0.15\%$ and $0.95\%$ for the Hubble constant $H_0$ and the constant EoS of dark energy $w$, respectively; for a two-parameter EoS parametrization of dark energy, the precision of $w_0$ is $2.04\%$ and the error of $w_a$ is 0.13. We conclude that 3G GW detectors would lead to breakthroughs in solving the Hubble tension and revealing the nature of dark energy provided that the EoS of NS is perfectly known.}

\maketitle
\section{Introduction} \label{sec:intro}
Allan Sandage pointed out in as early as 1970 that the two key numbers that should be precisely measured in cosmology are the Hubble constant ($H_0$) and the deceleration parameter ($q_0$) \cite{sandage1970cosmology}. This is undoubtedly a deep insight. The measurement of $q_0$ using type Ia supernovae has led to the discovery of the accelerating expansion of the universe \cite{SupernovaSearchTeam:1998fmf,SupernovaCosmologyProject:1998vns}, which is usually explained through assuming an exotic component with a negative pressure that is currently dominating the evolution of the universe, dubbed dark energy. However, the nature of dark energy is still an enigma, which might be solved if the equation of state (EoS) of dark energy is precisely measured. Therefore, in the current cosmology, an important mission is to precisely measure the Hubble constant and the EoS of dark energy.

The Hubble constant is a direct measure of the current expansion rate of the universe and it is a key cosmological parameter because all the cosmic properties relevant to the absolute scale, such as the age and size of the universe as well as some physical processes like the growth of cosmic structure and the nucleosynthesis of the light isotopes, are determined by it. Lots of efforts have been made to measure the Hubble constant in the past near one century, and currently the precisions of the measurements have reached (for indirect measurement) or nearly reached (for direct measurement) $1\%$. However, the values of the Hubble constant inferred from the observations of the $Planck$ cosmic microwave background (CMB) anisotropies (a $0.8\%$ measurement, assuming a flat $\Lambda$CDM cosmology) \cite{Planck:2018vyg} and determined from the Cepheid-supernova distance ladder (a $1.4\%$ measurement) \cite{Riess:2021jrx} are highly inconsistent, with the tension reaching 5$\sigma$ significance \cite{Riess:2021jrx}, which is the so-called ``Hubble tension'', greatly challenging the current cosmology. Recently, the Hubble tension has been widely discussed in the literature (see, e.g., refs. \cite{Riess:2019qba,Verde:2019ivm,Guo:2018ans,Gao:2021xnk,DiValentino:2021izs,Abdalla:2022yfr,cai:2020,Cai:2021wgv,Yang:2018euj,DiValentino:2019jae,DiValentino:2019ffd,DiValentino:2020zio,Liu:2019awo,Zhang:2019cww,Ding:2019mmw,Li:2020tds,Wang:2021kxc,Vagnozzi:2021tjv,Vagnozzi:2021gjh,Guo:2019dui,Vagnozzi:2019ezj,Feng:2019jqa,Lin:2020jcb,Hryczuk:2020jhi,Gao:2022ahg,Zhao:2022yiv}).

For the measurement of the EoS of dark energy (assuming a constant $w$), the CMB+BAO+SN data have given the tightest constraint to date, with the constraint precision being about $3\%$ \cite{Brout:2022vxf}. However, such a measurement precision is still far away from revealing the fundamental nature of dark energy.

The gravitational wave (GW) standard siren observation opened a new window into studying the expansion history of the universe, which recently has been widely discussed \cite{Holz:2005df,Dalal:2006qt,Cutler:2009qv,Zhao:2010sz,Cai:2016sby,Cai:2017aea,Cai:2017plb,Zhang:2019ylr,Zhao:2018gwk,Du:2018tia,Cai:2018rzd,Yang:2019bpr,Yang:2019vni,Bachega:2019fki,Chang:2019xcb,He:2019dhl,Zhao:2019gyk,Wang:2021srv,Qi:2021iic,Jin:2021pcv,Zhu:2021bpp,deSouza:2021xtg,Wang:2022oou,Wu:2022dgy,Jin:2022tdf,Song:2022siz,Hou:2022rvk,Chen:2020dyt,Gray:2019ksv,Califano:2022syd,Wang:2022rvf,Dhani:2022ulg,Colgain:2022xcz,Cao:2021zpf,Leandro:2021qlc,Ye:2021klk,Chen:2020zoq,Mitra:2020vzq,Hogg:2020ktc,Nunes:2020rmr,Borhanian:2020vyr,Jin:2023zhi,Jin:2023sfc} (see e.g., ref.~\cite{Bian:2021ini} for a recent review).
The luminosity distance of a GW source could be directly obtained through the analysis of the GW waveform without the external calibration, which is called a standard siren \cite{Schutz:1986gp,Holz:2005df}.
Furthermore, some GW sources, e.g., binary neutron star (BNS) mergers, could produce rich electromagnetic (EM) signals, such as kilonovae, short $\gamma$-ray bursts and their afterglows, referred to as EM counterparts of GWs \cite{Li:1998bw,Nakar:2007yr}. GWs together with their EM counterparts, if can both be observed, realize the multi-messenger observations, which are usually referred to as bright sirens, and could enable us to establish the absolute luminosity distance--redshift relation to constrain $H_0$ and $w(z)$.
In fact, the only multi-messenger observation, GW170817 from the first detected BNS merger, has given an independent constraint on the Hubble constant, at a $14\%$ precision \cite{LIGOScientific:2017adf}.
Limited by the sensitivities of the current GW detectors (LIGO and Virgo), only two BNS mergers are detected during the first, second, and third observing runs \cite{LIGOScientific:2018mvr,LIGOScientific:2020ibl,LIGOScientific:2021djp} and only a few BNS mergers will be detected during the fourth observing run \cite{KAGRA:2013rdx}. The Hubble constant would be measured to a $2\%$ precision using about 50 GW170817-like standard sirens \cite{Chen:2017rfc}, which may help resolve the Hubble tension.
However, in order to effectively measure the EoS of dark energy, considering only the second-generation GW detectors is definitely not hopeful, and one would has to resort to the next-generation detectors.

In the next decades, the third-generation (3G) ground-based GW detectors, the Cosmic Explorer (CE) \cite{LIGOScientific:2016wof} and the Einstein Telescope (ET) \cite{Punturo:2010zz}, will have a powerful detection sensitivity, one order of magnitude improved over the current GW detectors \cite{Evans:2021gyd}.
%which can be improved by one order of magnitude compared to the current GW detectors \cite{Evans:2021gyd}.
A series of forecasts \cite{Safarzadeh:2019pis,Belgacem:2019tbw,Yu:2021nvx} show that the 3G GW detectors will detect $\mathcal{O}(10^5)$ BNS mergers per year.
However, even so, only a small part of them (around $0.1\%$) have the detectable EM counterparts \cite{Yu:2021nvx}, and thus the application of bright sirens in cosmology has strong limitations.
Furthermore, it has been shown in some works \cite{Bian:2021ini,Jin:2021pcv,Zhang:2019loq,Jin:2020hmc} that solely using GW bright sirens cannot tightly constrain the EoS of dark energy and the main role they play in cosmology is to help break the cosmological parameter degeneracies of other traditional observations.

In 2012, Messenger and Read \cite{Messenger:2011gi} proposed that the distance--redshift relation can be established by solely using
the GW observations based on the 3G GW detectors (note here that the GW standard sirens without EM counterparts are called ``dark sirens''). This is because the tidal deformations of NSs in the BNS systems could be measured using the 3G GW detectors. The tidal deformations depend on the EoS of NS and could provide an additional contribution to the phase evolution of the GW waveform. The additional phase is related to the intrinsic mass and thus it can be used to break the degeneracy between mass and redshift, which leads to the redshift of BNS being measured.

Recently, using the GW dark sirens (tidal measurements of BNS) to measure the cosmological parameters has been discussed in the literature \cite{DelPozzo:2015bna,Wang:2020xwn,Chatterjee:2021xrm,Ghosh:2022muc}. Del Pozzo \emph{et al.} \cite{DelPozzo:2015bna} forecasted the ability of GW dark sirens from ET to measure the cosmological parameters. Wang \emph{et al.} \cite{Wang:2020xwn} forecasted the constraints on $w_0$ and $w_a$ (with other cosmological parameters fixed) using GW dark sirens from the 3G ground-based GW network. Chatterjee \emph{et al.} \cite{Chatterjee:2021xrm} forecasted the measurements of the Hubble constant using GW dark sirens from O5, Voyager, and CE. Ghosh \emph{et al.} \cite{Ghosh:2022muc} forecasted the measurements of the Hubble constant using GW dark sirens from CE.
%Actually, utilizing the GW dark sirens of

Previous works focused on the ability to measure the Hubble constant using the mock dark siren data of a single 3G GW observatory. Actually, in the next decades, the 3G GW detectors are expected to form a ground-based detector network, thereby improving the ability to detect BNS mergers. In this work, we wish to answer the question of whether the Hubble constant and the EoS of dark energy can both be precisely measured using only the GW dark sirens detected by the 3G ground-based GW detector network. We make cosmological analysis using four cases of 3G GW observations, single ET, single CE, CE-CE network (one CE in the US and another one in Australia, abbreviated as 2CE hereafter), and ET-CE-CE network (abbreviated as ET2CE hereafter).
We note that GW dark sirens using the cross-correlation of binary black hole mergers and galaxy catalogs \cite{Nair:2018ign,DES:2019ccw,DES:2020nay,LIGOScientific:2019zcs,Gray:2019ksv,Yu:2020vyy,LIGOScientific:2021aug,Palmese:2021mjm,Wang:2022oou,Song:2022siz} are not discussed here. For cosmological models, we consider the $w$CDM and $w_0w_a$CDM models to make cosmological analysis. We employ $\Lambda$CDM as the fiducial model to generate the mock GW dark siren data, with the fiducial values of cosmological parameters being set to the constraint results of $Planck$ 2018 TT,TE,EE+lowE \cite{Planck:2018vyg}.

This work is organized as follows. In section \ref{subsec1}, we briefly introduce the merger rate of BNS adopted in this work. In section \ref{subsec2}, we introduce the method of calculating the detection number of BNS merger events. In section \ref{subsec3}, we introduce the method of simulating GW dark sirens. In section \ref{subsec4}, we briefly introduce the method of constraining cosmological parameters. In section \ref{sec:results}, we show the constraint results. In section \ref{sec:discussion}, we compare the constraint results of this work with those of other future observations. The conclusion is given in section \ref{sec:conclusion}.

\section{Methodology}\label{sec:Method}

\subsection{BNS merger rate}\label{subsec1}
The merger rate as a function of redshift in the observer frame is given by
\begin{equation}
R_{\rm o}(z)=\frac{R_{\rm m}(z)}{1+z} \frac{d V(z)}{d z},
\end{equation}
where the $(1+z)$ term arises from converting the source-frame time to the observer-frame time and $dV/dz$ is the comoving volume element. $R_{\rm m}$ is the merger rate in the source frame, which is related to the cosmic star formation rate,
\begin{equation}
R_{\rm m}\left(z_{\rm m}\right)=\int_{z_{\rm m}}^{\infty} d z_{\rm f} \frac{d t_{\rm f}}{d z_{\rm f}} R_{\rm sf}\left(z_{\rm f}\right) P\left(t_{\rm d}\right),
\end{equation}
where ${d t}/{d z}=-[(1+z)E(z)H_0]^{-1}$. The time-delay distribution $P(t_{\rm d})$ is the probability density that the binary system formed at time $t_{\rm f}$ (or redshift $z_{\rm f}$) and merged at time $t_{\rm m}$ (or redshift $z_{\rm m}$) with $t_{\rm d}=t_{\rm m}-t_{\rm f}$. $R_{\rm sf}$ is the cosmic star formation rate. We assume the Madau-Dickinson star formation rate \cite{Madau:2014bja}, with an exponential time delay (an e-fold time of 100 Myr) \cite{Vitale:2018yhm}. Meanwhile, we adopt the local comoving merger rate to be 320 $\rm Gpc^{-3}~yr^{-1}$ that is the estimated median rate based on the O3a observation \cite{LIGOScientific:2020kqk}.

\subsection{The detection number of BNS merger events}\label{subsec2}
\begin{figure}
\includegraphics[width=0.6\textwidth]{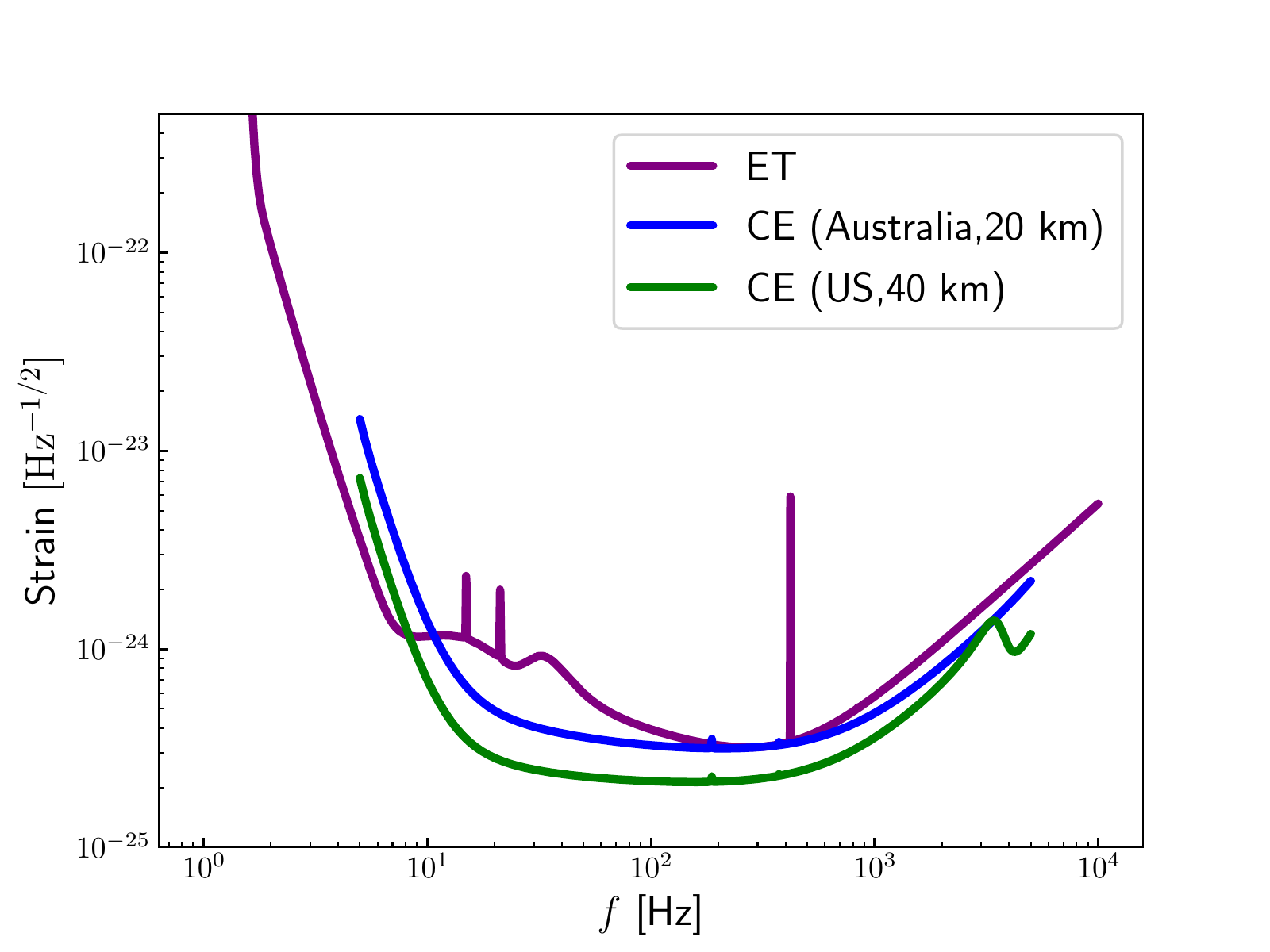}
\centering
\caption{\label{fig1} {Sensitivity curves of ET \cite{ETcurve-web} and CE \cite{CEcurve-web} used in this work. Note that we consider one 40 km-arm-length CE in the US and one 20 km-arm-length CE in Australia.}}
\end{figure}

Then we calculate the detection number of BNS merger events. Here we adopt the signal-to-noise ratio (SNR) threshold to be 8 for a detection. The combined SNR for the detection network of $N$ detectors is $\rho=\sqrt{\sum\limits_{a=1}^{N}(\rho_{a})^2}$, where $\rho_a$ is the SNR of the $a$th detector and is given by
\begin{equation}
\rho_a^{2}=\frac{5}{6} \frac{\left(G \mathcal{M}_{\rm chirp}\right)^{5 / 3}}{c^{3} \pi^{4 / 3} d_{\rm L}^{2}(z)} \int_{f_{\rm lower}}^{f_{\rm upper }} d f \frac{\mathcal{F}_{a}^{2}f^{-7 / 3}}{S_{a}(f)},
\end{equation}
where $\mathcal{M}_{\rm chirp}=(1+z)\eta^{3/5}M$ is the observed chirp mass, $M=m_1+m_2$ is the total mass of a binary system with the component masses $m_1$ and $m_2$, $\eta=m_1 m_2/(m_1+m_2)^2$ is the symmetric mass ratio, $d_{\rm L}$ is the luminosity distance to the GW source, $G$ is the gravitational constant, $c$ is the speed of light, $f_{\rm lower}$ is the lower cutoff frequency ($f_{\rm lower}=1$ Hz for ET and $f_{\rm lower}=5$ Hz for CE), $f_{\rm upper}=2/(6^{3/2}2\pi M_{\rm obs})$ is the frequency at the last stable orbit with $M_{\rm obs}=(m_1+m_2)(1+z)$ \cite{Zhao:2010sz}, and $S_{a}(f)$ represents the one-side noise power spectral density (PSD) of the $a$-th GW detector. We wish to note that in this work we consider the waveform in the inspiralling stage for a non-spinning BNS system. Hence, the upper cutoff frequency is chosen as the frequency at the last stable orbit. If the inspiral-merger-ringdown waveform is considered, a higher SNR is expected to provide better measurements on $d_{\rm L}$, thus providing better constraints on cosmological parameters. We adopt the PSD of ET from ref.~\cite{ETcurve-web} and that of CE from ref.~\cite{CEcurve-web} (we consider one 40 km-arm-length CE in the US and one 20 km-arm-length CE in Australia and the noise curves are shown in figure \ref{fig1}).
$\mathcal{F}_{a}=\sqrt{\frac{\left(1+\cos ^{2} \iota\right)^{2}}{4} F_{+, a}^{2}+\cos ^{2} \iota F_{\times, a}^{2}}$ characterizes the detector response, where $\iota$ is the inclination angle between the binary's orbital angular momentum and the line of sight, $F_{+, a}$ and $F_{\times, a}$ are the $a$th detector's antenna response functions of two GW polarizations, $+$ and $\times$. Note that we adopt the NS's mass distribution from the analysis of the latest O3 observation (figure 7 of ref.~\cite{KAGRA:2021duu}).

The detector response functions are related to the source location, the polarization angle, the detector's location (latitude $\varphi$, longitude $\lambda$, the angle between the interferometer arms $\zeta$, and the orientation angle $\gamma$ of the detector's arms measured counter-clockwise from East of the Earth to the bisector of the interferometer arms). In table \ref{tab1}, we list the coordinates of the GW observatories considered in this work. Some other 3G GW detectors' location candidates could be found in ref.~\cite{Gossan:2021eqe}. We note that the 3G GW detectors have much wider frequency range than the current ground-based GW detectors. The time of BNS merger is a function of $f^{-8/3}$. Thus, the BNS signals can be found in the 3G GW band for hours, even for several days, and in the 2G GW band for tens of minutes \cite{Zhao:2017cbb}. Therefore, the time dependence in the detector response functions is considered in this work. The forms of $F_+$ and $F_{\times}$ are related to these parameters, as detailedly described in refs.~\cite{Jaranowski:1998qm,Arnaud:2001my,Schutz:2011tw}.

\begin{table}
\centering
\setlength\tabcolsep{7pt}
\renewcommand{\arraystretch}{1.5}
\caption{\label{tab1} {The coordinates of the GW observatories \cite{lalsuite,Ashton:2018jfp,di2021seismological,Borhanian:2020ypi} considered in this work.}}
\begin{tabular}{ccccc}
\hline
{GW detector}  & {$\varphi\ [\mathrm{deg}]$} & {$\lambda\ [\mathrm{deg}]$} & {$\gamma\ [\mathrm{deg}]$} & {$\zeta\ [\mathrm{deg}]$} \\
\hline
{Cosmic Explorer, USA} &  {$43.827$} & {$-112.825$} & {$45.000$} & {90.000} \\
{Einstein Telescope, Europe} &  {$40.443$} & {$9.457$} & {$0.000$} & {60.000} \\
{Cosmic Explorer, Australia} &  {$-34.000$} & {$145.000$} & {$90.000$} & {90.000} \\
\hline
\end{tabular}
\end{table}

\begin{figure}
\includegraphics[width=0.6\textwidth]{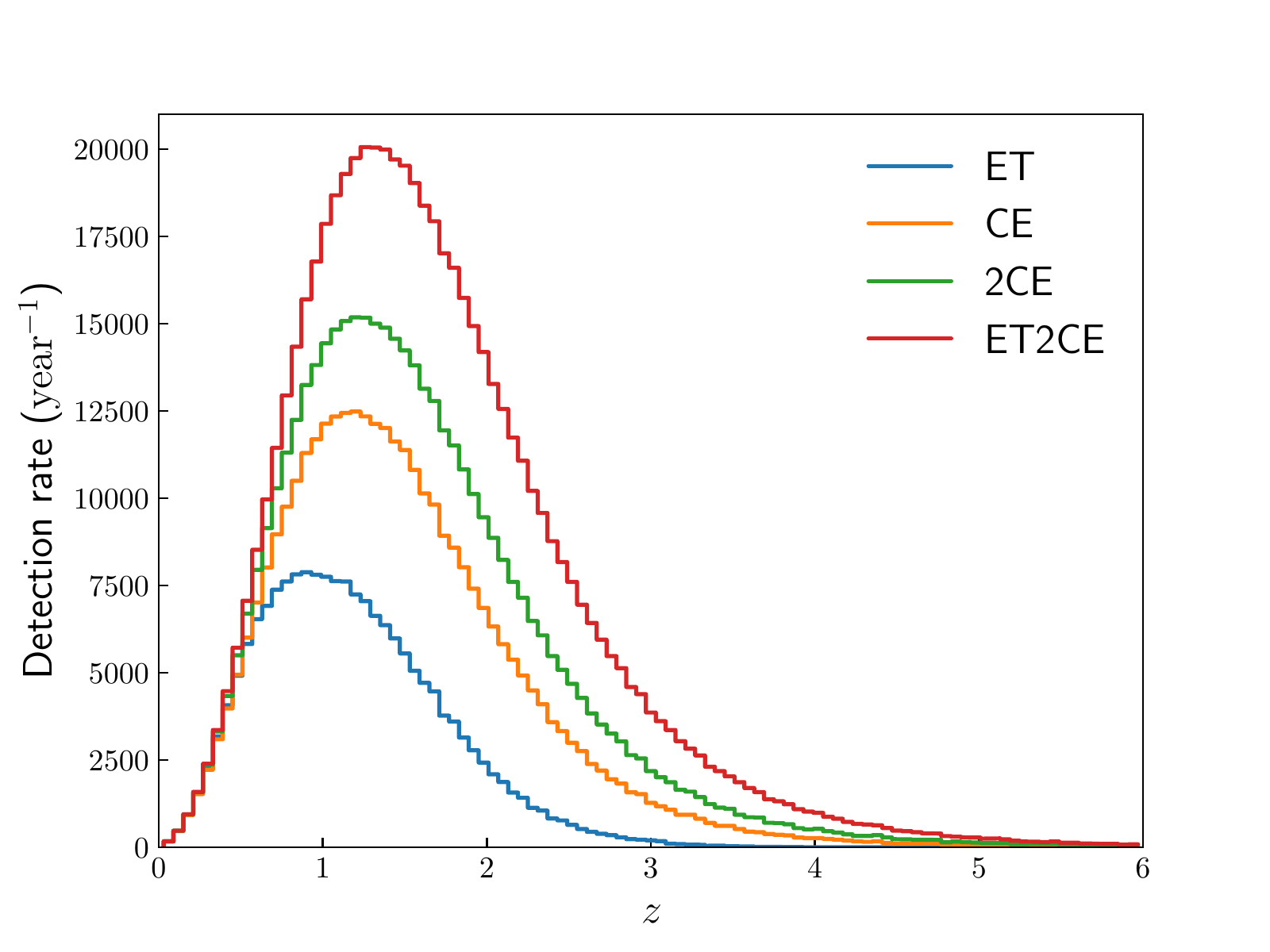}
\centering
\caption{\label{fig2}The expected detection rate as a function of redshift for ET, CE, 2CE, and ET2CE.}
\end{figure}

According to our calculation, we obtain the numbers of the GW events detected by ET (545763), CE (1017618), the 2CE network (1317183), and the ET2CE network (1815501), assuming a three-year observation. We see that the number of GW events detected by CE is more than twice of ET. Compared to the single CE observatory, the number of GW events detected by the 2CE network almost doubles. Meanwhile, the number of GW events detected by the ET2CE network is slightly more than that of the 2CE network. In figure~\ref{fig2}, we show the expected detection rate as a function of redshift for ET, CE, 2CE, and ET2CE.

\subsection{GW dark sirens}\label{subsec3}
Under the stationary phase approximation \cite{Zhang:2017srh}, the Fourier transform of the time-domain waveform for a detector network (with $N_d$ detectors) is given by \cite{Zhao:2017cbb}
\begin{equation}
\tilde{\boldsymbol{h}}(f)=e^{-i \bf\Phi}\hat{\boldsymbol{h}}(f),
\end{equation}
{with $\hat{\boldsymbol{h}}$ given by}
\begin{equation}
{\hat{\boldsymbol{h}}(f)=\left[\frac{{h}_1 (f)}{\sqrt{S_1(f)}},\frac{{h}_2 (f)}{\sqrt{S_2(f)}},\cdots,\frac{{h}_{N_d} (f)}{\sqrt{S_{N_d}(f)}}\right],}
\end{equation}
where $\bf\Phi$ is the $N_d\times N_d$ diagonal matrix with $\Phi_{ab}=2\pi f\delta_{ab}(\textbf{n}\cdot\textbf{r}_a)$, $\textbf{n}$ is the propagation direction of a GW, and $\textbf{r}_a$ is the location of the $a$th detector. Here $S_{a}(f)$ is PSD of the $a$-th detector.
We consider the waveform in the inspiralling stage for a non-spinning BNS system. We adopt the restricted Post-Newtonian (PN) approximation and calculate the waveform to the 3.5PN order \cite{Sathyaprakash:2009xs,Blanchet:2004bb},
\begin{align}
h_a(f)=\mathcal{A}_a&f^{-7/6}\exp \{i\big(2 \pi f t_{\rm c}-\pi / 4-2\psi_{\rm c} +2 \Psi(f / 2)-\varphi_{(2,0),a}+\Phi_{\text {tidal}}(f)\big)\},
\end{align}
where the Fourier amplitude $\mathcal{A}_a$ is given by
\begin{align}
\mathcal{A}_a=&~~\frac{1}{d_{\rm L}}\sqrt{\big(F_{+,a}(1+\cos^2\iota)\big)^2+(2F_{\times,a}\cos\iota)^2}
\times \sqrt{5\pi/96}\pi^{-7/6}\mathcal{M}_{\rm chirp}^{5/6}.
\end{align}
Here $\psi_{\rm c}$ is the coalescence phase and the detailed forms of $\Psi(f)$ and $\varphi_{(2,0),a}$ can be found in refs.~\cite{Sathyaprakash:2009xs,Blanchet:2004bb}.
Note that we employ the stationary phase approximation to obtain the frequency-domain expression of GW waveform, in which the time $t$ in $F_+$, $F_{\times}$, and $\Phi_{ij}$ is replaced by $t_{\rm f}=t_{\rm c}-(5 / 256) \mathcal{M}_{\rm chirp}^{-5 / 3}(\pi f)^{-8 / 3}$ \cite{Maggiore:2007ulw,Zhao:2017cbb}, with $t_{\rm c}$ the coalescence time.

The tidal contribution to the GW phase from a BNS system is given by \cite{Messenger:2011gi}
\begin{equation}
\Phi_{\text {tidal}}(f)= \sum_{j=1}^2 \frac{3 \lambda_{j}}{128 \eta M^{5}}\bigg[-\frac{24}{\chi_{j}}\big(1+\frac{11 \eta}{\chi_{j}}\big)(\pi Mf)^{5/3}-\frac{5}{28 \chi_{j}}\big(3179-919 \chi_{j}-2286 \chi_{j}^{2}+260 \chi_{j}^{3}\big) (\pi Mf)^{7/3}\bigg],
\end{equation}
where the sum is comprised of the components of the binary system, $\chi_j=m_j/(m_1+m_2)$, and $\lambda_j$ is the tidal deformation, which is related to the EoS of NS and characterizes changes of the quadruple of NS. It is comparable in magnitude with the 3.5PN phasing term for NSs \cite{Messenger:2011gi}. Due to the fact that EoS of NS is still unknown, we choose SLy \cite{Douchin:2001sv} as the fiducial EoS of NS that is also consistent with the current observation \cite{LIGOScientific:2017vwq}. {In fact, the consideration of different EoSs of NS may lead to different constraint results. In ref.~\cite{Messenger:2011gi}, Messenger and Read found that for the three typical NS's EoSs, the redshift measurement errors obtained by using the tidal measurement method can differ by several times (a stiffer EoS gives better redshift measurement). In ref.~\cite{Wang:2020xwn}, Wang \emph{et al.} used different EoSs of NS and focused on constraining dark energy by fixing other cosmological parameters (a stiffer EoS of NS gives better constraints). Hence, in this work, in order to make the cosmological analysis representative, we choose a medium EoS of NS, namely SLy.}

In this work, following ref.~\cite{Wang:2020xwn}, in order to parameterize the EoS of NS, we express the $\lambda$--$m$ relation as a linear function of the NS mass in the range of $1.2~M_{\odot}-2~M_{\odot}$ as
\begin{equation}
\lambda=Bm+C,
\end{equation}
where $B$ and $C$ are tidal effect parameters. For the fitting of SLy, we obtain $B=-2.14$ and $C=4.64$, where the goodness-of-fit value is 0.999. Therefore, the influence of the fitting errors on the results is almost negligible.

Then we can use the Fisher information matrix to calculate the measurement errors of the GW parameters.
For a set of parameters $\boldsymbol{\theta}$, the Fisher information matrix of a network including $N_d$ independent GW detectors is given by
\begin{equation}
{F}_{ij}=\left(\frac{\partial \tilde{\boldsymbol{h}}}{\partial \theta_i}\Bigg\vert \frac{\partial \tilde{\boldsymbol{h}}}{\partial \theta_j}\right),
\end{equation}
where $\theta_i$ denotes twelve parameters ($d_{\rm L}$, $z$, $\mathcal{M}_{\rm chirp}$, $\eta$, $t_{\rm c}$, $\psi_{\rm c}$, $\iota$, $\delta$, $\alpha$, $\psi$, $B$, $C$) for a given BNS system.
The inner product is defined as
\begin{equation}
\left( \zeta\vert \xi\right)=2 \int_{f_{\text {lower }}}^{f_{\mathrm{upper}}} \{{\zeta}(f) {\xi}^{*}(f)+{\zeta}^{*}(f) {\xi}(f)\} df,
\end{equation}
where $*$ represents complex conjugate. Here we numerically calculate the partial derivatives $\partial\tilde{h}(f)/\partial \theta_i$ by $[\tilde{h}(f,\theta_{i}+d\theta_i)-\tilde{h}(f,\theta_{i})]/d\theta_i$, with $d\theta_i=10^{-n}$. In order to verify the robustness of our code, we compare our Fisher matrix with that obtained from the state-of-the-art code $\tt GWFAST$ \cite{Iacovelli:2022mbg}. We find that the results are consistent after optimizing the value of $n$ to ensure the convergence of the derivative. The Fisher information matrix can then be used to calculate the measurement errors of the GW parameters
\begin{equation}
\Delta \theta_i=\sqrt{(F^{-1})_{ii}}. \label{error}
\end{equation}
The instrumental error is $\sigma_{\rm d_{\rm L}}^{\rm inst}=\Delta d_{\rm L}=\sqrt{(F^{-1})_{11}}$.
The measurement of $d_{\rm L}$ is also affected by the weak lensing and we adopt the form \cite{Tamanini:2016zlh,Speri:2020hwc,Hirata:2010ba}
\begin{equation}
\sigma_{d_{\rm L}}^{\rm lens}(z)=\left[1-\frac{0.3}{\pi / 2} \arctan \left(z / 0.073\right)\right]\times d_{\rm L}(z)\times 0.066\bigg[\frac{1-(1+z)^{-0.25}}{0.25}\bigg]^{1.8}.
\end{equation}
The total error of $d_{\rm L}$ is given by
\begin{align}
\sigma_{d_{\rm L}}=\sqrt{(\sigma_{d_{\rm L}}^{\rm inst})^2+(\sigma_{d_{\rm L}}^{\rm lens})^2}.
\end{align}

\begin{figure}
\includegraphics[width=1.1\textwidth]{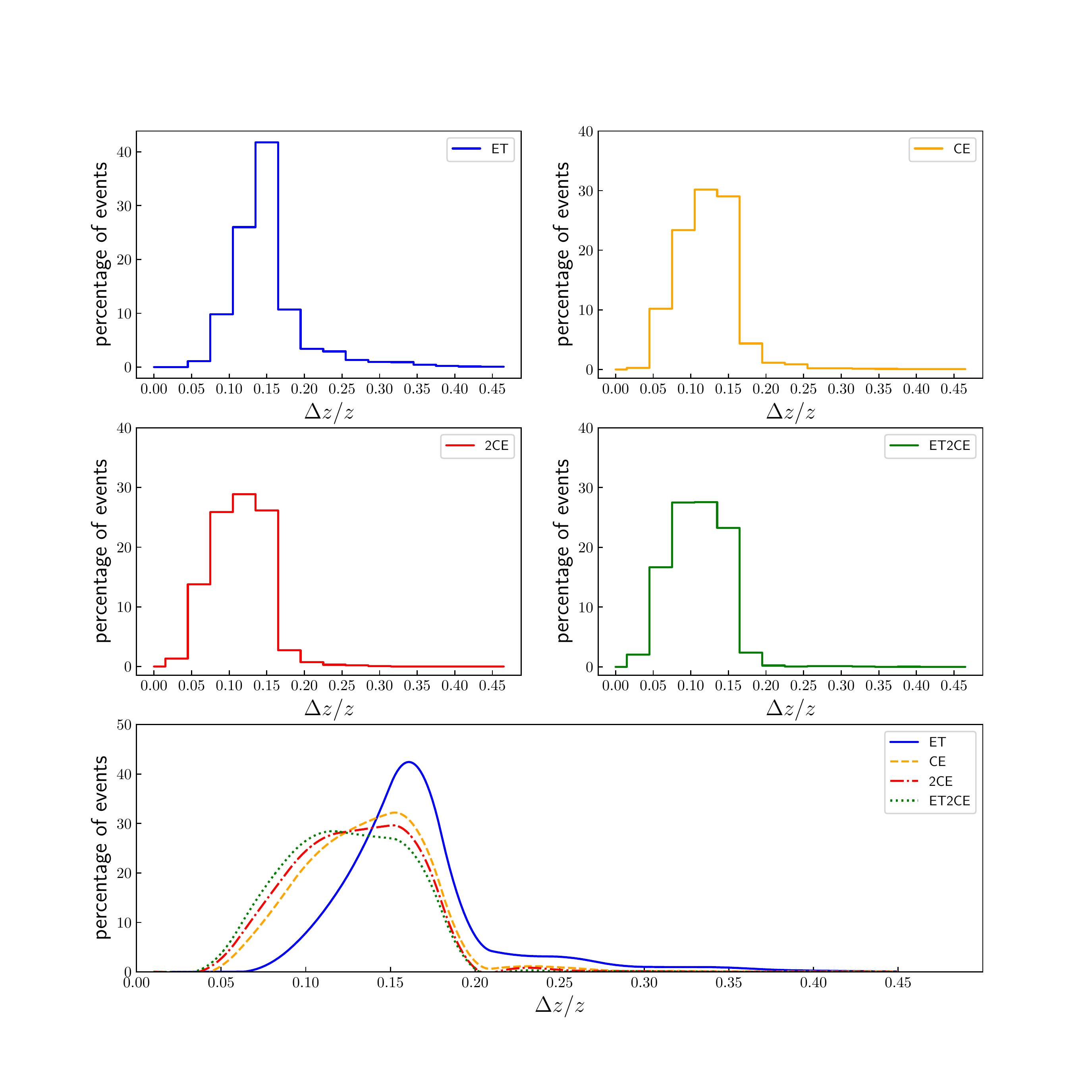}
\centering
\caption{\label{fig3} {Percentage of $\Delta z/z$ distribution of GW events detected by ET, CE, 2CE, and ET2CE.
In the upper four panels, we show the percentages of $\Delta z/z$ distributions for ET, CE, 2CE, and ET2CE in blue, orange, red, and green lines, respectively.
In the lower panel, we use the interpolation method to fit the upper four panels to show the distribution of $\Delta z/z$ intuitively.}}
\end{figure}

Throughout this work, we simulate the BNS mergers with random binary orientations and sky directions. The sky direction, binary inclination, polarization angle, and the coalescence phase are randomly chosen in the ranges of ${\rm cos}\,\delta\in [-1,1]$, $\alpha\in [0,360^\circ]$, ${\rm cos}\,\iota\in [-1,1]$, $\psi\in [0,360^\circ]$, and $\psi_{\rm c}\in [0,360^\circ]$. {For the NS's mass, we adopt the mass distribution based on the analysis of the latest O3 observation (figure 7 of ref.~\cite{KAGRA:2021duu}).} Without loss of generality, the merger time is chosen to $t_{\rm c}=0$ in our analysis. The simulated GW dark sirens satisfy the redshift distributions corresponding to ET, CE, 2CE, and ET2CE shown in figure~\ref{fig2}.

When using BNS tidal deformation measurements to obtain redshift information, the GW tidal phases need to be precisely measured to obtain the source-frame mass information (related to $B$ and $C$). Combining with the observed chirp mass information obtained from the measurements of observed GW frequency $f$ and its time derivative $\dot{f}$, the redshift information could be obtained by breaking the parameter degeneracies between mass and redshift. In our analysis, we consider the correlation between the twelve parameters to calculate the measurement errors of redshifts. Using eq.~(\ref{error}), we can obtain the measurement errors of redshifts, $\Delta z=\sqrt{(F^{-1})_{22}}$.

In the upper and middle panels of figure~\ref{fig3}, we show the distributions of $\Delta z/z$ for ET, CE, 2CE, and ET2CE. In order to be more intuitive, in the lower panel of figure~\ref{fig3}, we use the interpolation method to fit the histograms above. We see that the measurement precisions of redshifts are mainly $5\%$--$20\%$. ET2CE gives the best redshift measurement, followed by 2CE, CE, and ET. The ET case shows evident difference compared with the other three cases, while the CE, 2CE, and ET2CE cases also have some slight differences.

\begin{figure}
\includegraphics[width=0.8\textwidth]{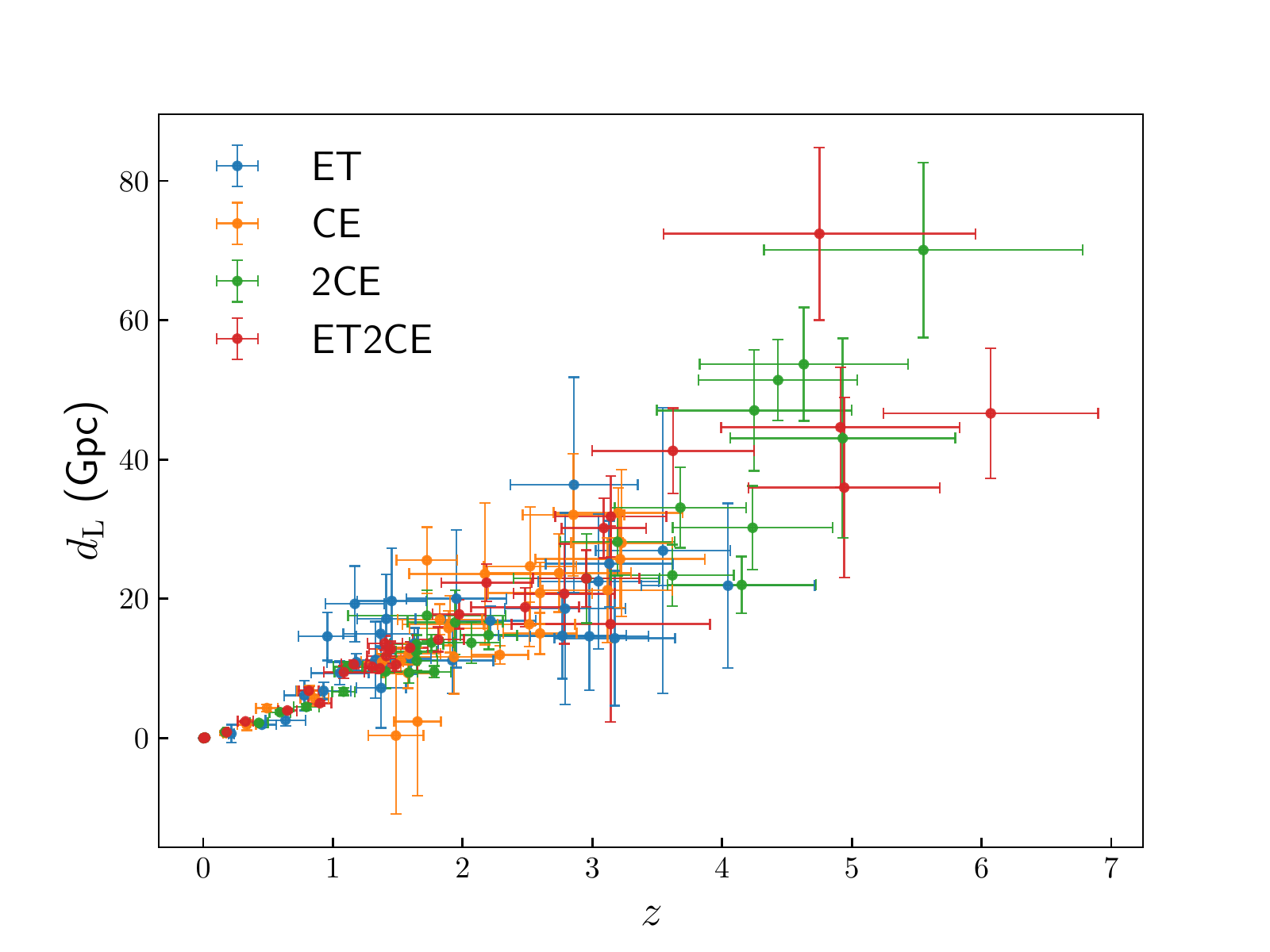}
\centering
\caption{\label{fig4} The simulated standard sirens observed by ET, CE, 2CE, and ET2CE. Here we show only 30 data points for each observation for simplicity. We actually simulate 545763, 1017618, 1317183, and 1815501 standard sirens for ET, CE, 2CE, and ET2CE.}
\end{figure}

In figure~\ref{fig4}, we show the simulated GW standard sirens observed by ET, CE, 2CE, and ET2CE. For simplicity, we show only 30 data points for each observation. To account for fluctuations in the measured values resulting from actual observations, we represent the standard siren data points with Gaussian randomization. Specifically, the central value is populated according to a Gaussian distribution with mean equal to the fiducial value and standard deviation equal to the measurement error.

\subsection{Method of constraining cosmological parameters}\label{subsec4}

Given a set of $N$ GW data $\vec{\mathcal{D}}$, the posterior probability distribution of the cosmological parameters $\vec{\Omega}$ is
\begin{equation}
p(\vec{\Omega}\vert\vec{\mathcal{D}})\propto\mathcal{L}(\vec{\mathcal{D}}\vert\vec{\Omega})p(\vec{\Omega}),
\end{equation}
where $p(\vec{\Omega})$ is the prior on cosmological parameters $\vec{\Omega}$. The likelihood function for the GW data $\mathcal{L}(\vec{\mathcal{D}}\vert\vec{\Omega})$ is given by
\begin{equation}
\mathcal{L}(\vec{\mathcal{D}}\vert\vec{\Omega})=\prod_{i}^{N} p(d_{\mathrm{L, obs}}^{i}\vert d_{\mathrm{L}}(z^{i},\vec{\Omega})) p(z_{\mathrm{obs}}^{i}\vert z^{i}) p(z^{i}), \label{likelihood}
\end{equation}
where $i$ represents the $i$-th GW event, $p(d_{\mathrm{L, obs}}^{i}\vert d_{\mathrm{L}}(z^{i},\vec{\Omega}))$ and $p(z_{\mathrm{obs}}^{i}\vert z^{i})$ are two Gaussian distributions with the standard deviations calculated from eq.~(\ref{error}), and $p(z^{i})$ is the redshift prior that is assumed to be uniform for simplicity. For the dark siren method, the redshifts $z$ of the GW events are not precisely measured. Therefore, the redshift $z$ should be marginalized out to obtain the constraints on cosmological parameters, i.e., eq.~(\ref{likelihood}) should be rewritten as
\begin{equation}
\mathcal{L}(\vec{\mathcal{D}}\vert\vec{\Omega})=\prod_{i}^{N} \int p(d_{\mathrm{L, obs}}^{i}\vert d_{\mathrm{L}}(z^{i},\vec{\Omega})) p(z_{\mathrm{obs}}^{i}\vert z^{i})\mathrm{d}z^{i}.
\end{equation}
Note that in the above equation $p(z^{i})$ is not explicitly shown because it is assumed to be a constant.

Here, we also note that the method described above is valid under two assumptions: (i) the measurements of $z$ and $d_{\rm L}$ are assumed to be nearly independent, with $d_{\rm L}$ measured from the GW amplitude and $z$ measured from the GW phase, and (ii) the detection probability (selection bias) does not depend on the observed values of $z$ and $d_{\rm L}$. In a more realistic approach, the observed values of $z$ and $d_{\rm L}$ may deviate from the simulated values. In such cases, it is necessary to use the observed data to calculate SNRs and the detection probability. If assumption (ii) is not satisfied, the Bayesian inference could be corrected by adding the redshift prior conditioned on detection, which is basically the detection rate as a function of redshift (see ref.~\cite{Ding:2018zrk} for an example).

\begin{table}[!htbp]
\caption{The absolute errors (1$\sigma$) and the relative errors of the cosmological parameters in the $w$CDM and $w_0w_a$CDM models by using the mock data of ET, CE, 2CE, and ET2CE.}
\label{tab:full}
\setlength\tabcolsep{15pt}
\renewcommand\arraystretch{1.5}
\begin{center}{\centerline{
\begin{tabular}{cccccc}
\hline
          Model       & Parameter &ET & CE & 2CE & ET2CE \\ \hline
 \multirow{6}{*}{$w$CDM}& $\sigma(\Omega_m)$& $0.0055$ & $0.0028$   & $0.0013$ & $0.0011$ \\
 &$\sigma(H_0)$ &$0.270$ &$0.187$ & $0.125$ & $0.103$ \\
  &$\sigma(w)$&$0.034$ &$0.020$  & $0.012$ & $0.010$ \\
 &$\varepsilon(\Omega_{\rm m})$&$1.75\%$ &$0.89\%$ & $0.43\%$ & $0.32\%$ \\
 &$\varepsilon(H_0)$&$0.40\%$ &$0.27\%$ & $0.18\%$ & $0.15\%$ \\
 &$\varepsilon(w)$&$3.39\%$ &$2.01\%$ & $1.19\%$ & $0.95\%$ \\ \hline
 \multirow{7}{*}{$w_0w_a$CDM}& $\sigma(\Omega_m)$ &$0.0255$ &$0.0113$  & $0.0059$ & $0.0046$\\
 &$\sigma(H_0)$&$0.366$ &$0.250$ & $0.185$ & $0.160$ \\
  &$\sigma(w_0)$&$0.050$ &$0.034$  & $0.024$ & $0.020$ \\
  &$\sigma(w_a)$&$0.504$ &$0.271$ & $0.157$ & $0.126$ \\
 &$\varepsilon(\Omega_{\rm m})$&$8.09\%$ &$3.56\%$ & $1.89\%$ & $1.48\%$ \\
 &$\varepsilon(H_0)$&$0.54\%$ &$0.37\%$ & $0.27\%$ & $0.24\%$ \\
 &$\varepsilon(w_0)$&$5.01\%$ &$3.42\%$ & $2.39\%$ & $2.04\%$ \\
 \hline
\end{tabular}}}
\end{center}
\end{table}

\section{Results} \label{sec:results} % and discussion

\begin{figure}
\includegraphics[width=7.6cm,height=7.6cm]{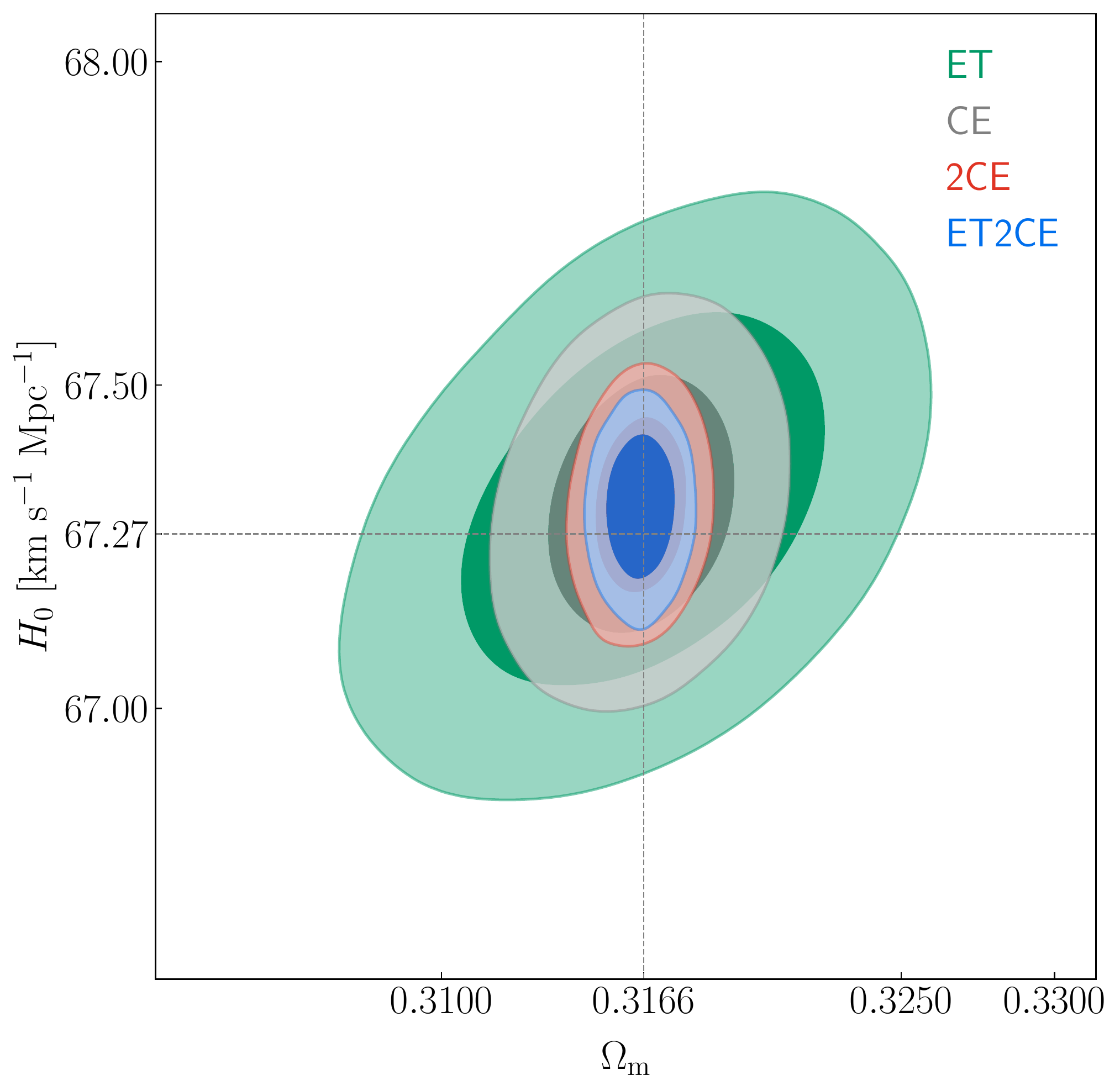}%scale=0.55
\includegraphics[width=7.6cm,height=7.6cm]{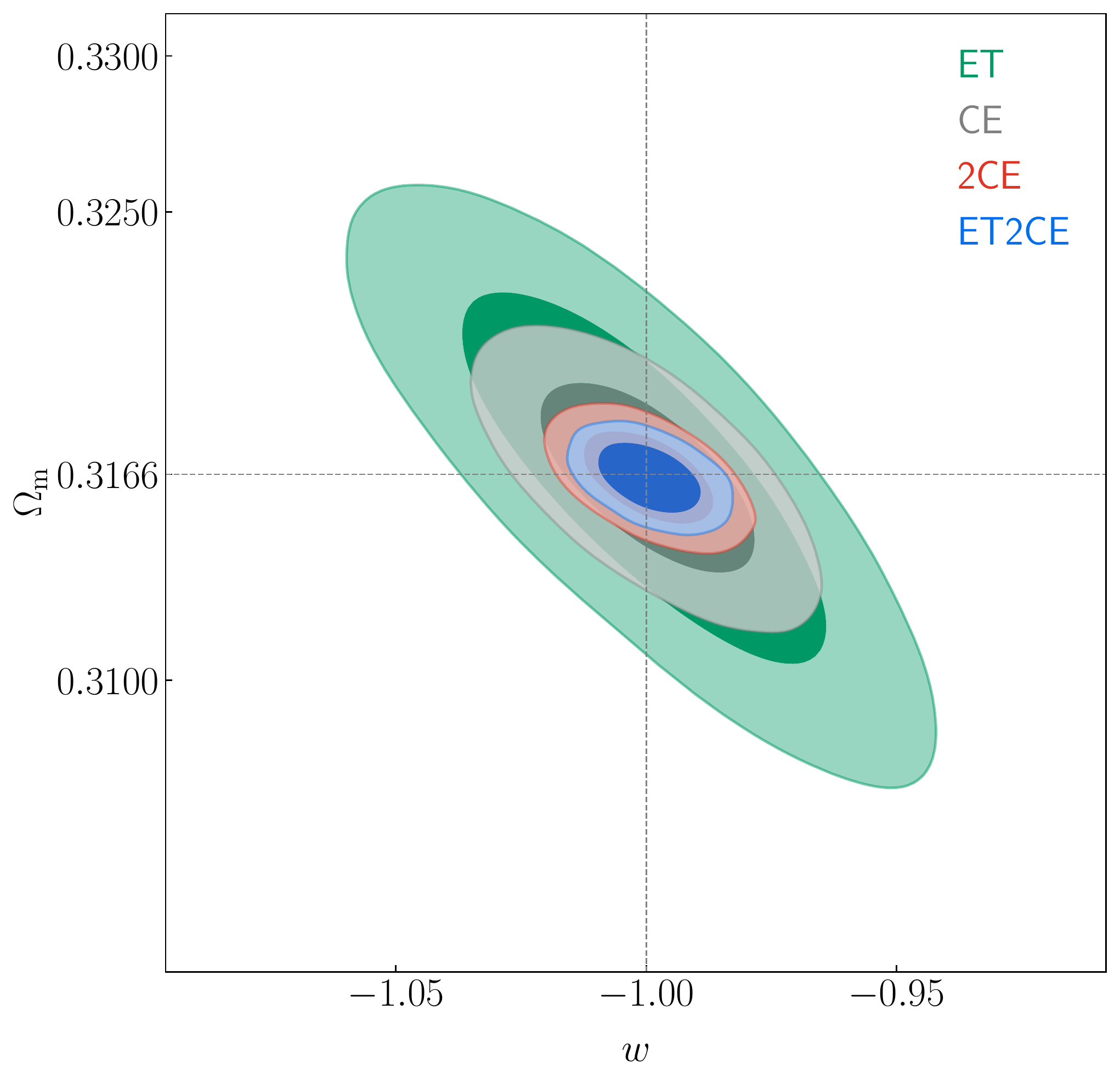}%scale=0.55
\centering
\caption{\label{fig5} Two-dimensional marginalized contours ($68.3\%$ and $95.4\%$ confidence level) in the $\Omega_{\rm m}$--$H_0$ and $w$--$\Omega_{\rm m}$ planes using the mock GW dark siren data of ET, CE, 2CE, and ET2CE. Here, the dotted lines indicate the fiducial values of cosmological parameters preset in the simulation.}
\end{figure}
\begin{figure}
\includegraphics[width=7.6cm,height=7.6cm]{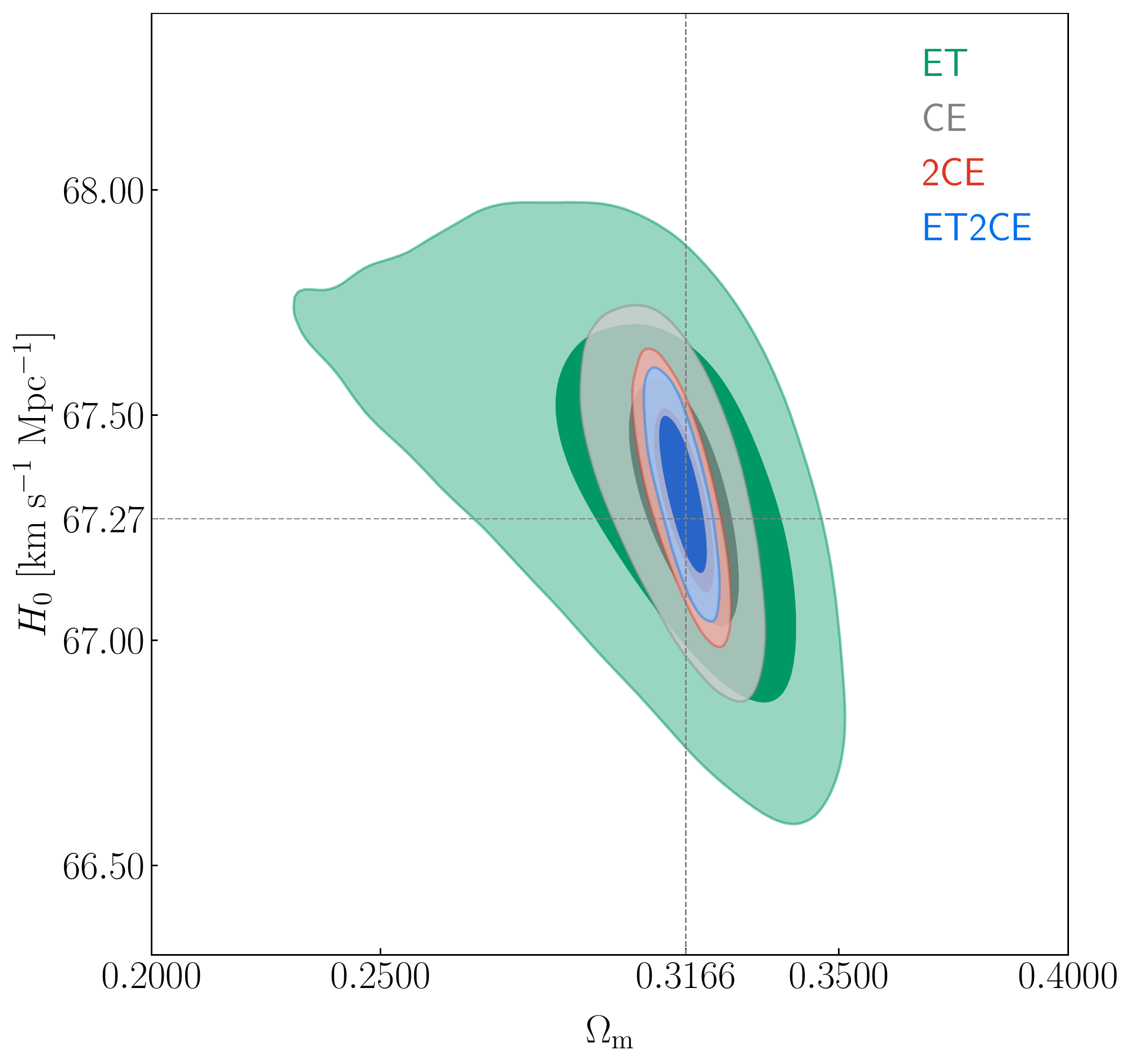}
\includegraphics[width=7.6cm,height=7.6cm]{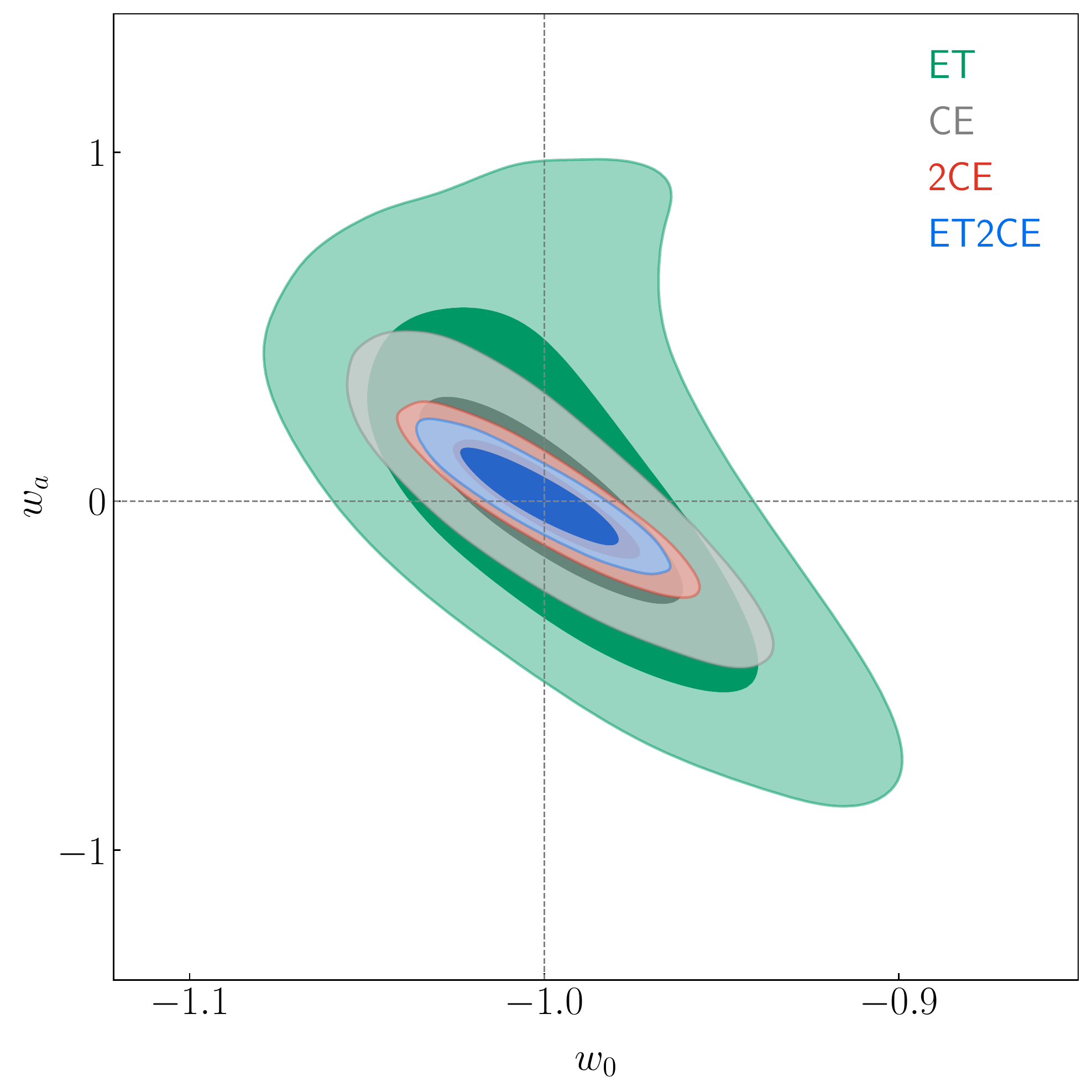}
\centering
\caption{\label{fig6} Two-dimensional marginalized contours ($68.3\%$ and $95.4\%$ confidence level) in the $\Omega_{\rm m}$--$H_0$ and $w_0$--$w_a$ planes using the mock GW dark siren data of ET, CE, 2CE, and ET2CE. Here, the dotted lines indicate the fiducial values of cosmological parameters preset in the simulation.}
\end{figure}

We use the simulated GW dark siren data from ET, CE, 2CE, and ET2CE and by performing the MCMC analysis \cite{Lewis:2002ah} we use the mock data of dark sirens to constrain the $w$CDM and $w_0w_a$CDM models. The constraint results are shown in figures~\ref{fig5} and \ref{fig6} and summarized in table~\ref{tab:full}. For a parameter $\xi$, we use $\sigma(\xi)$ and $\varepsilon(\xi)$ to represent its absolute (1$\sigma$) and relative errors, respectively, with $\varepsilon(\xi)$ defined as $\sigma(\xi)/\xi$.

We first take a look at the $w$CDM model. In the left panel of figure~\ref{fig5}, we show the constraints in the $\Omega_{\rm m}$--$H_0$ plane. We can clearly see that GWs place quite tight constraints on the parameters. Owing to the fact that the detection sensitivity of CE is better than that of ET, ET shows the worst constraint results among the four cases. However, even so, ET gives $\sigma(\Omega_{\rm m})=0.0055$, $\sigma(H_0)=0.270$, and $\sigma(w)=0.034$ which are comparable with the results of $\sigma(\Omega_{\rm m})=0.0065$, $\sigma(H_0)=0.660$, and $\sigma(w)=0.028$ by the latest Planck TT,TE,EE+lowE+BAO+Pantheon data \cite{Brout:2022vxf}.
Meanwhile, ET and CE show positive correlation between $\Omega_{\rm m}$ and $H_0$, while there is almost no correlation between $\Omega_{\rm m}$ and $H_0$ when using the mock GW dark siren data of 2CE and ET2CE. However, the obvious anti-correlation is obtained when using the traditional EM cosmological probes.
In the right panel of figure~\ref{fig5}, we show the constraints in the $w$--$\Omega_{\rm m}$ plane. We find that the 2CE network can give quite tight constraints on the cosmological parameters with the constraint precisions of $\Omega_{\rm m}$, $H_0$, and $w$ being $0.43\%$, $0.18\%$, and $1.19\%$, respectively. So, even for the EoS of dark energy, the standard of precision cosmology is nearly achieved. Furthermore, we find that using the ET2CE mock data, the constraint precisions of $\Omega_{\rm m}$, $H_0$, and $w$ are $0.32\%$, $0.15\%$, and $0.95\%$, respectively, all meeting the standard of precision cosmology.

In figure~\ref{fig6}, we show the constraint results of the $w_0w_a$CDM model.
Among the four cases, ET gives the worst constraint results. CE gives much better constraints than ET. 2CE gives similar constraint results with ET2CE. Quantitatively,
ET gives $\sigma(w_0)=0.050$ and $\sigma(w_a)=0.504$, comparable with the results of $\sigma(w_0)=0.064$ and $\sigma(w_a)=0.300$ by the latest Planck TT,TE,EE+lowE+BAO+Pantheon data \cite{Brout:2022vxf}. Meanwhile, the constraint results of ET are also basically consistent with those given in ref.~\cite{DelPozzo:2015bna}. For the case of using CE, the constraint results are much better than those of ET, also better than those of the latest Planck TT,TE,EE+lowE+BAO+Pantheon data. What's more, when considering 2CE and ET2CE, the constraint precisions of $w_0$ are $2.39\%$ (2CE) and $2.04\%$ (ET2CE). For $w_a$, ET2CE gives $\sigma(w_a)=0.126$.

\section{Discussion}\label{sec:discussion}

In section \ref{sec:results}, we compare the constraint results of the GW dark sirens (tidal measurements) with the current cosmological observations, from which the potential of the GW dark sirens can be clearly seen. Since third-generation GW detectors will begin observation in the 2030s, it is also necessary to show the comparisons with other cosmological observations in the same period.

\subsection{{Comparison with future other GW standard siren observations}}
In the 3G GW detector network era, lots of bright sirens could also be used to study the cosmic expansion history. Belgacem \emph{et al.} \cite{Belgacem:2019tbw} used the mock bright siren data (based on the 10-year observation of ET2CE and THESEUS) to measure dark energy and obtained $\sigma(w)=0.034$ for the optimistic case of observing EM counterparts. When combining the mock bright siren data of ET2CE with the current CMB+BAO+SN data, the constraint errors could be reduced to $\sigma(w)=0.013$ that is comparable with the constraint results of 2CE in this work. For the case of the $w_0w_a$CDM model, the combination of the mock bright siren data and the CMB+BAO+SN data gives $\sigma(w_0)=0.025$ and $\sigma(w_a)=0.137$ \cite{Belgacem:2019tbw} that are also comparable with those of 2CE in this work.

In the same period of the 3G GW detectors, the space-based GW detectors will begin detecting GWs generated by the massive or supermassive black hole binaries (SMBHBs). Wang \emph{et al.} \cite{Wang:2021srv} used the mock bright siren data of the LISA-Taiji network to measure dark energy. Zhu \emph{et al.} \cite{Zhu:2021aat} used the mock bright siren data of LISA-TianQin network to measure dark energy. However, owing to the fact that the number of the detectable bright sirens of SMBHBs are small (about $\mathcal{O}(10)$) within 5-year observation, the constraint results are all weaker than those of this work.

\subsection{{Comparison with future EM surveys}}
In the next decades, the fourth-generation dark energy surveys will also precisely measure dark energy \cite{Pierel:2020tav,Euclid:2021qvm}. Pierel \emph{et al.} \cite{Pierel:2020tav} forecasted the ability of cosmological parameter estimation using strongly lensed SN from the Roman Space Telescope (previously WFIRST). The CMB+BAO+SN+UltraDeep lensed SN data give $\sigma(w)=0.01$ in the $w$CDM model, and $\sigma(w_0)=0.02$ and $\sigma(w_a)=0.14$ in the $w_0w_a$CDM model, comparable with those of 2CE in this work. Recently, Euclid Collaboration \cite{Euclid:2021qvm} forecasted the constraint results of Euclid. The pessimistic case of Euclid (see ref.~\cite{Euclid:2021qvm} for more details) gives $\sigma(w_0)=0.038$ and $\sigma(w_a)=0.14$ that are comparable with those of 2CE. If considering the combination of CMB-S4 and Euclid (pessimistic), the constraint results are worse than those of ET2CE. While considering the optimistic case of Euclid (see ref.~\cite{Euclid:2021qvm} for more details), the absolute errors are reduced to $\sigma(w_0)=0.021$ and $\sigma(w_a)=0.073$ that are comparable with those of ET2CE. If considering the combination of CMB-S4 and Euclid (optimistic), the constraint results are also comparable with those of ET2CE.

\subsection{{Comparison with future other cosmological probes}}
In the future, neutral hydrogen 21 cm intensity mapping (IM) will become an important tool to explore the nature of dark energy \cite{Bull:2014rha,Xu:2014bya,SKA:2018ckk,Zhang:2021yof,Wu:2021vfz,Wu:2022jkf,Zhang:2023gaz}. Bull \emph{et al.} \cite{Bull:2014rha} obtained $\sigma(w_0)=0.035$ and $\sigma(w_a)=0.163$ by using the SKA1-MID+MeerKAT data, which are comparable with those of 2CE. Xu \emph{et al.} \cite{Xu:2014bya} obtained $\sigma(w_0)=0.082$ and $\sigma(w_a)=0.21$ by using the full-scale Tianlai data, which are comparable with those of ET. Wu \emph{et al.} \cite{Wu:2022dgy} forecasted cosmological parameters using the combination of four promising cosmological probes, 21 cm IM, GW, fast radio burst, and strong gravitational lensing, and obtained $\sigma(w)=0.020$, comparable with that of CE.

\section{Conclusion} \label{sec:conclusion}
In the next decades, the third-generation GW detectors will detect large amounts of ($\mathcal{O}(10^5)$ per year) GW events generated by the BNS mergers. By measuring the additional tidal deformation-phase term, the redshifts of the GW sources could be obtained. Hence, the cosmic expansion history can be investigated using solely GW observations. In this work, based on the three-year observation, we simulate such GW dark sirens of ET, CE, 2CE, and ET2CE to perform cosmological analysis. Note that the assumption of the EoS of NS being perfectly known is adopted in this work. We wish to investigate whether the Hubble constant and the EoS of dark energy can both be precisely measured using only GW dark sirens detected by the 3G GW observations.

We find that ET gives the worst constraints among the four cases. Even so, ET gives $\sigma(w)=0.034$ that is comparable with the latest constraint result of Planck TT,TE,EE+lowE+ BAO + Pantheon. Meanwhile, the constraint precision of $w$ using the mock 2CE data is close to $1\%$, which almost meets the standard of precision cosmology. The 3G GW detector network has a strong capability to constrain the EoS parameter of dark energy $w$, with a precision of 0.95\%. For the Hubble constant, the single ET observatory gives quite tight constraint with the precision being $0.40\%$. Using the ET2CE network, we obtain the constraint precision of $H_0$ being $0.15\%$.

We also discuss the case of the two-parameter dark-energy model. We find that ET gives comparable constraint results with those of the latest Planck TT,TE,EE+lowE+ BAO + Pantheon data. CE gives much better constraint results than those of ET. ET2CE gives slightly better constraint results than 2CE. Using ET2CE, we obtain the constraint precision of $w_0$ being $2.04\%$. Moreover, ET2CE gives $\sigma(w_a)=0.126$ that is much better than the result of the latest Planck TT,TE,EE+lowE+BAO+Pantheon data.

In addition, we discuss the comparison of the constraint results of GW dark sirens (tidal measurements) with those of the same-period other cosmological observations. We find that GW dark sirens show strong ability to measure the Hubble constant and dark energy. The constraints on the Hubble constant and the EoS of dark energy using GW dark sirens of ET2CE are comparable with those of the combination of the fourth-generation dark energy surveys (optimistic case of Euclid) and CMB-S4.

Our results show that the precision cosmology can be achieved by solely using the GW dark sirens observed by the 3G GW detectors. It can be expected that the 3G GW detectors would play a key role in helping solve the Hubble tension and reveal the fundamental nature of dark energy.

\acknowledgments
We thank Ji-Yu Song, Jiming Yu, Liang-Gui Zhu, Arnab Dhani, Ssohrab Borhanian, Peng-Ju Wu, Ling-Feng Wang, Tao Han, Jing-Zhao Qi, Bo Wang, and Rui-Qi Zhu for helpful discussions. This work was supported by the National SKA Program of China (Grants Nos. 2022SKA0110200 and 2022SKA0110203), the National Natural Science Foundation of China (Grants Nos. 11975072, 11875102, and 11835009), and the 111 Project (Grant No. B16009).

\bibliography{gwtidal}{}
\bibliographystyle{JHEP}

\end{document}